\documentclass[aip,jcp,reprint,longbibliography]{revtex4-2}

\usepackage{amsmath}
\usepackage{amssymb}
\usepackage{amsfonts}
\usepackage{amsthm}
\usepackage{bm}
\usepackage{physics}
\usepackage{graphicx}
\usepackage{xcolor}
\usepackage{hyperref}
\usepackage{mathtools}
\usepackage{enumitem}
\usepackage{booktabs}
\usepackage{multirow}
\usepackage{dcolumn}
\usepackage{float}
\usepackage{subcaption}
\usepackage{tikz}
\usepackage{braket}

\ifdefined\AnnotatedVersion
  \definecolor{RoneBlue}{RGB}{0,10,110}
  \definecolor{RtwoGreen}{RGB}{0,10,65}
  \newcommand{\Rone}[1]{\textcolor{RoneBlue}{#1}}
  \newcommand{\Rtwo}[1]{\textcolor{RoneBlue}{#1}}
  \newcommand{\Rconcept}[1]{\textcolor[rgb]{0.545,0,0}{#1}}
\else
  \newcommand{\Rone}[1]{#1}
  \newcommand{\Rtwo}[1]{#1}
  \newcommand{\Rconcept}[1]{#1}
\fi

\begin{document}


\title{\Rconcept{Liouvillian Geometry of Multidimensional
Spectra: Pathway Transport and Local Response Deformation}}

\author{Eric R. Bittner}
\email{ebittner@central.uh.edu}
\affiliation{Department of Physics, University of Houston, Houston, Texas 77204, USA}
\affiliation{Institut Courtois \& D\'epartement de physique, Universit\'e de Montr\'eal, 1375 Avenue Th\'er\`ese-Lavoie-Roux, Montréal, Qu\'ebec H2V~0B3, Canada}

\author{Carlos~Silva-Acu\~na}
\email{carlos.silva@umontreal.ca}
\affiliation{Institut Courtois \& D\'epartement de physique, Universit\'e de Montr\'eal, 1375 Avenue Th\'er\`ese-Lavoie-Roux, Montréal, Qu\'ebec H2V~0B3, Canada}
\affiliation{School of Chemistry and Biochemistry, Georgia Institute of Technology, 901 Atlantic Drive, Atlanta, GA~30332, United~States}
\affiliation{Departamento de F\'isica Aplicada, Centro de Investigaci\'on y de Estudios Avanzados del Instituto Polit\'ecnico Nacional (CINVESTAV), 97310 M\'erida, Yucat\'an, M\'exico}

\author{Hao Li}
\affiliation{Institut Courtois \& D\'epartement de physique, Universit\'e de Montr\'eal, 1375 Avenue Th\'er\`ese-Lavoie-Roux, Montréal, Qu\'ebec H2V~0B3, Canada}

\date{}

\begin{abstract}
Liouville pathways provide the conceptual foundation for interpreting
multidimensional spectroscopies but, in open quantum systems, need not
remain independent under the reduced dynamics. Environmental interactions
can transport amplitude between pathway sectors during free evolution,
producing measurable changes in the nonlinear response. We develop a
differential framework for describing this transport under parametric
deformation of the open-system dynamics. When the environment selects a
pointer basis distinct from the observational basis, the resulting basis
incompatibility mixes Liouville pathways and produces characteristic
spectral distortions. Using a Duhamel expansion of the Liouvillian
propagator, we derive local response derivatives that reconstruct this
pathway redistribution. For a two-level model with a rotated dephasing
basis, a second-order expansion about a single reference point retains
the dominant spectral structure over a broad range of basis
\Rconcept{misalignment, while the global error grows to approximately
$37\%$ at maximal misalignment. These results show how local response
information captures the spectroscopic consequences of Liouvillian
deformation and establish pathway transport as a complementary organizing
principle for multidimensional spectroscopy.}
\end{abstract}

\maketitle


\section{Introduction}
\label{sec:intro}

Multidimensional spectroscopies provide some of the most 
powerful tools available for probing quantum dynamics in 
condensed-phase and molecular systems. By correlating 
excitation, waiting, and detection intervals, techniques 
such as two-dimensional electronic spectroscopy (2DES), 
two-dimensional infrared spectroscopy (2DIR), and nonlinear 
magnetic resonance experiments resolve couplings, coherence 
transfer pathways, population relaxation, and environmental 
fluctuations that are inaccessible to linear response 
measurements~\cite{Mukamel1995,Jonas2003,Cho2008,Fuller2015,
Brixner2004,Hamm2011}. Central to the interpretation of 
these experiments is the Liouville-space pathway picture, 
introduced and systematized by Mukamel and 
co-workers~\cite{Mukamel1995,Mukamel2000}, in which the 
measured signal is expressed as a sum of contributions 
generated by a sequence of field-matter interactions and 
subsequent propagation under an effective system 
Liouvillian. Double-sided Feynman diagrams provide a 
compact bookkeeping for these pathways and have become 
the standard language of nonlinear 
spectroscopy~\cite{Mukamel1995,Jonas2003,Brixner2005,
Cho2008,Engel2007}.

The pathway description has proven extraordinarily
successful. It has been applied to study vibrational
and electronic couplings~\cite{Brixner2005,Khalil2003},
energy transfer and relaxation dynamics in photosynthetic
complexes~\cite{Engel2007,Fleming2009,Schlau-Cohen2011},
coherence and dephasing in molecular
aggregates~\cite{Cho2008,Mukamel2000}, and the interplay
of vibrational and electronic coherence across a range
of systems~\cite{Jonas2018,Cao2020}. Despite this success,
the pathway picture rests on an implicit assumption: the
basis used to define the pathways also provides an
approximately independent and physically transparent
representation of the dynamics between optical interactions.

In practice, quantum systems are never isolated.
Interactions with the surrounding environment continuously
reshape the evolving density matrix, generating relaxation,
dephasing, coherence transfer, and population
redistribution~\cite{Redfield1957,Lindblad1976,Breuer2002}.
As a result, the observed spectrum need not correspond
precisely to the idealized pathway decomposition generated
by a Hamiltonian model. When resolved in a fixed
pathway basis, the resulting open-system dynamics may
redistribute amplitude among nominally distinct pathway
components, producing spectral distortions whose
microscopic origin is often difficult to identify. Standard 
treatments address this through the secular Redfield approximation~\cite{Redfield1957,Bloch1957},
which decouples population and coherence evolution and approximately
preserves the conventional pathway structure, or through more complete
treatments such as the
hierarchical equations of motion
(HEOM)~\cite{Tanimura1989,Ishizaki2009}, which capture
non-Markovian effects but often obscure a simple
pathway-based physical interpretation.

The failure of the observational basis to coincide with
the bath-selected basis is not merely a technical
complication; it has a precise physical meaning. Following
Zurek’s theory of environment-induced superselection
(einselection)~\cite{Zurek1981,Zurek1982,Zurek2003}, the
environment dynamically selects a preferred or pointer
basis consisting of the states most robust under
environmental monitoring. When the pointer basis and
Hamiltonian eigenbasis coincide, this source of pathway
mixing vanishes and the conventional pathway decomposition
is recovered. When they do not, the environment acts in a
basis incompatible with the pathway decomposition and
continuously mixes pathways that the idealized model treats
as independent. This basis incompatibility provides the
physical mechanism for pathway transport and a natural
parameterization of the resulting deformation of the
spectroscopic response.

This observation raises a natural inverse problem. Given 
a model Hamiltonian and its associated multidimensional 
spectrum, can one determine how environmental interactions 
modify the underlying pathway structure? Can the difference 
between an idealized spectrum and an observed spectrum be 
interpreted in terms of a transport process acting in 
Liouville space? Answering these questions would provide 
a route toward quantifying the dynamical influence of the 
environment directly from spectroscopic observables, 
without treating environmental effects solely through 
phenomenological broadening or relaxation 
parameters~\cite{Mukamel1995,Tokmakoff2000}.

Geometric concepts have increasingly shaped our
understanding of quantum dynamics and response. Berry
phases and holonomies arising from cyclic evolution in
parameter space leave observable signatures in
spectroscopic signals~\cite{Berry1984,Aharonov1987}, and
quantum geometric tensors provide a unified framework
connecting state distinguishability, adiabatic transport,
and response~\cite{ProvostVallee1980,Zanardi2007}. More
recently, the stationary-state response of open quantum
systems has been shown to admit a geometric decomposition
into symmetric and antisymmetric sectors~\cite{Bittner:2026aa,
Bittner:2026ab}, with the antisymmetric sector governing
nonreciprocal response and geometric work. In the present
setting, basis incompatibility between the Hamiltonian
eigenbasis and the environment-selected pointer basis
provides a physically transparent deformation through which
this response geometry can be probed spectroscopically.

The present work extends this perspective to nonlinear
spectroscopy by developing a differential description of
pathway transport under parametric deformation of the
open-system dynamics. Within a minimal rotated
pointer-basis model, the deformation is quantified by an
angle $\theta$ measuring the misalignment between the
Hamiltonian eigenbasis and the environment-selected pointer
basis. \Rone{For the spectroscopic calculations considered
here, the parameter-dependent environmental interaction is
represented by pure dephasing in the rotated pointer basis,
while a fixed population-relaxation channel is retained as
part of the reference dynamics. The resulting spectra are
evaluated from the total rephasing third-order response at
a fixed waiting time, $t_2$.}
Using the Duhamel identity, we construct local response
derivatives that describe how the nonlinear spectrum
changes as the Liouvillian is deformed. This provides a
route toward \emph{geometric inverse spectroscopy}, in
which environmental pathway transport is inferred from
multidimensional spectroscopic observables.

\Rone{An important feature of the present construction is
that the Duhamel expansion is intrinsically local in model
space: response derivatives evaluated at a reference
Liouvillian are used to reconstruct the spectrum at finite
parameter displacement. As demonstrated below, a
second-order expansion about the aligned reference model
at $\theta=0$ retains the dominant spectral structure even
for displacements approaching $\theta=\pi/2$, a substantially
larger useful range than might be anticipated for such a
low-order local expansion. Quantitative errors nevertheless
grow with increasing displacement from the reference model.
This behavior both demonstrates the information encoded in
local response derivatives and exposes the limitation of
finite-order extrapolation from a single reference point.
It provides a natural motivation for subsequent
autodifferentiable transport approaches in which local
response information is reevaluated incrementally along a
trajectory through model space.}

The remainder of this paper is organized as follows.
Section~\ref{sec:II} establishes pathway transport as an
intrinsic feature of open-system dynamics and identifies
basis incompatibility as a physical mechanism for pathway
mixing. Section~\ref{sec:III} develops the differential
description of parametric Liouvillian transport and relates
it to the geometry of the underlying state–generator model
manifold. Section~\ref{sec:IV} introduces the minimal
rotated pointer-basis model, derives the local Liouvillian
expansion, and constructs the pathway-transport response
using the Duhamel identity.
\Rone{Section~\ref{sec:V} presents the resulting
spectroscopic signatures and assesses the range over which
a second-order local expansion reconstructs the nonlinear
response.}
Section~\ref{sec:VI} discusses the broader implications
for response transport, geometric inverse spectroscopy,
and extensions beyond single-point local reconstruction.
\section{Pathway Transport in Open Quantum Systems}
\label{sec:II}

Conventional interpretations of multidimensional 
spectroscopic signals rest on the assignment of Liouville 
pathways generated by sequences of field--matter 
interactions. This framework has proven remarkably 
successful for describing coherence transfer, relaxation 
dynamics, and environmental fluctuations. Implicit in 
this construction, however, is the assumption that the 
pathway basis remains fixed throughout the experiment 
--- that the basis used to define the pathways also 
provides an adequate representation of the dynamics 
between optical interactions.

Open quantum systems present a different situation. 
\Rone{When the dissipative and observational structures are
not simultaneously diagonalizable,} pathways are not static objects but
dynamical entities continuously reshaped by environmental interactions. 
The environment acts during every interval of free 
evolution, \Rone{and may redistribute} amplitude among nominally 
distinct pathways and generate coherence transfer 
and spectral distortions that cannot be understood 
solely in terms of the original pathway decomposition. 
The measured signal reflects the cumulative action of 
both optical and environmental transport.

To make this concrete, consider any system whose 
Hamiltonian eigenstates are delocalized superpositions 
of the states the environment monitors. During each 
free-evolution interval, the environment acts to 
suppress coherences in its preferred basis while 
the Hamiltonian generates coherences in its own 
eigenbasis. Neither wins outright. The result is a 
continuous redistribution of amplitude among the 
pathways that the idealized diagram treats as 
independent. The observed spectrum is not the 
idealized spectrum; it is the idealized spectrum 
after being acted on by this transport process.

\Rone{The minimal model studied below realizes this situation
through pure dephasing in a pointer basis rotated relative
to the Hamiltonian eigenbasis, together with a fixed
population-relaxation channel. In the aligned limit, the
dephasing contribution suppresses coherence without
coupling population and coherence sectors in the
Hamiltonian basis. Finite basis misalignment makes the
same dissipative operator non-diagonal in that basis and
thereby induces pathway mixing, while the population-
relaxation dynamics remain fixed.}

The dynamics are governed by a Liouvillian that 
separates naturally into three contributions,
\begin{align}
\mathcal{L}(t)
=
\mathcal{L}_H
+\mathcal{V}(t)
+
\mathcal{D}_{\rm bath},
\label{eq:Ldecomp}
\end{align}
where $\mathcal{L}_H\rho = -\frac{i}{\hbar}[H,\rho]$
generates coherent evolution, $\mathcal{V}(t)$ represents
the external field interactions that generate and probe the
spectroscopic pathways, and $\mathcal{D}_{\rm bath}$
collects the dissipative contributions from the environment.
The field interactions define the sequence of pathway
sectors sampled by the nonlinear response, while transport
during the intervening free-evolution intervals arises from
the generally incompatible action of $\mathcal{L}_H$ and
$\mathcal{D}_{\rm bath}$.

\Rone{More generally, we regard the reduced dynamical model
as depending on a set of external or structural parameters
$\boldsymbol{\lambda}=(\lambda^1,\ldots,\lambda^d)$, where
the index $\mu=1,\ldots,d$ labels coordinates in this
parameter space.  These parameters may characterize, for
example, Hamiltonian couplings, dissipative rates, or the
relative orientation of Hamiltonian and bath-selected bases.
The spectral reconstruction in Sec.~\ref{sec:V} follows a
one-dimensional slice with $\lambda^1=\theta$ and fixed
$\gamma_\phi=0.1$. The two-control geometry and full-domain
convergence study also vary $\gamma_\phi$.}

The physical origin of that incompatibility is the 
existence of two distinct preferred bases. The 
\emph{observational basis} is defined by the 
eigenstates of $\mathcal{L}_H$ --- the Hamiltonian 
eigenbasis in which the pathway diagram is drawn. 
The \emph{pointer basis} consists of the states most 
robust under environmental monitoring, selected 
dynamically through system--environment interactions 
following Zurek's theory of environment-induced 
superselection~\cite{Zurek1981,Zurek2003}. In 
Liouville space, these two structures are associated 
respectively with the eigenstructure of 
$\mathcal{L}_H$ and that selected by 
$\mathcal{D}_{\rm bath}$.

We emphasize that this distinction is made at the level of the
\emph{reduced open-system description}. For the combined system and
environment, the total dynamics may be represented as unitary evolution
on an enlarged Hilbert space, in which case no fundamental separation
between coherent and dissipative dynamics is required. After the
environmental degrees of freedom are eliminated, however, the effective
reduced generator contains coherent and dissipative structures that need
not be simultaneously diagonalizable. The basis incompatibility discussed
here therefore does not assert an absolute distinction between microscopic
coherent and dissipative dynamics. Rather, it characterizes the reduced
Liouvillian relative to the observational basis in which the nonlinear
spectroscopic pathways are resolved.

Pathway transport emerges when the coherent and
dissipative structures are incompatible,
\begin{align}
[\mathcal{L}_H,\mathcal{D}_{\rm bath}] \neq 0,
\label{eq:commutator}
\end{align}
so that the Hamiltonian-defined pathway basis and the
bath-selected pointer basis cannot be simultaneously
diagonalized. During each free-evolution interval, this
incompatibility allows the dissipative dynamics to
redistribute amplitude among sectors that are distinct in
the Hamiltonian-defined pathway decomposition. When the
two structures are compatible, this source of pathway
mixing vanishes and the conventional pathway decomposition
is recovered.

Within this reduced description, the incompatibility is
well defined: once the subsystem and effective Liouvillian
are specified, whether the coherent and dissipative
generators commute is independent of their representation.
Different microscopic system–bath models may therefore
give rise to the same reduced pathway-transport structure.

Multidimensional spectroscopy thus becomes a problem
of transport on a network of pathways. The objective
is not to identify every microscopic environmental
process individually, but to characterize how
environmental interactions reshape the pathway
structure itself. The relevant quantities are the
resulting changes in the spectroscopic response, rather
than the individual bath modes or spectral densities
that produce them. This reframing provides the foundation
for a differential description of pathway transport under
parametric deformation of the reduced dynamics, developed
in the following section.

\section{Parametric Transport of Liouvillian Response}
\label{sec:III}

Pathway transport can be formulated differentially by
considering a family of reduced dynamical models
$\mathcal{L}(\boldsymbol{\lambda})$ parameterized by the
model-space coordinates introduced in
Sec.~\ref{sec:II}. An infinitesimal displacement
$d\lambda^\mu$ changes both the generator of the
open-system dynamics and the propagators entering the
nonlinear response. The resulting derivatives provide a
local description of how spectroscopic pathways and their
associated response functions change under deformation of
the physical model.

It is useful to distinguish this local differential
construction from the geometric connection used to compare
models at different points of parameter space. The
Liouvillian derivative $\partial_\mu\mathcal{L}$ measures
the local deformation of the dynamical generator, while
derivatives of the propagator follow from the Duhamel
identity.
These objects generate the local response
expansion developed below, but they are not themselves a
geometric connection. A geometric connection requires
additional structure on the manifold of physical models.
In our related work~\cite{Bittner:2026ab},
this structure is constructed from parameterized
state--generator pairs, which define the model manifold
on which neighboring dynamical descriptions may be
compared geometrically.


\subsection{Parametric Variation of Liouvillian Eigenoperators}
\label{sec:IIIA}

Consider a family of reduced dynamical generators
$\mathcal{L}(\boldsymbol{\lambda})$ depending on the
model-space coordinates $\lambda^\mu$ introduced in
Sec.~\ref{sec:II}. These parameters may represent
Hamiltonian couplings, dissipative rates, external fields,
or structural properties of the environment. The
corresponding Liouvillian eigenproblem is
\begin{align}
\mathcal{L}(\boldsymbol{\lambda})
\Phi_n(\boldsymbol{\lambda})
=
\Lambda_n(\boldsymbol{\lambda})
\Phi_n(\boldsymbol{\lambda}),
\label{eq:LiouvillianEigenproblem}
\end{align}
where $\Lambda_n$ are the Liouvillian eigenvalues and
$\Phi_n$ denote the associated right eigenoperators.
For a non-Hermitian Liouvillian, we introduce the
corresponding left eigenoperators satisfying
\begin{align}
\widetilde{\Phi}_m^\dagger
\mathcal{L}
=
\Lambda_m
\widetilde{\Phi}_m^\dagger,
\end{align}
together with the biorthogonal normalization
\begin{align}
\langle\langle
\widetilde{\Phi}_m
|
\Phi_n
\rangle\rangle = 
\delta_{mn},
\end{align}
where
$\langle\langle A|B\rangle\rangle
=\mathrm{Tr}(A^\dagger B)$
denotes the Hilbert–Schmidt inner product.

As the model parameters are varied, both the eigenvalues
and eigenoperators may change. These variations have
distinct physical consequences. Changes in the
eigenvalues modify relaxation rates, dephasing rates, and
oscillation frequencies. Changes in the eigenoperators,
by contrast, modify the Liouville-space modes through
which the dynamics are resolved and can therefore mix
components that are distinct in a fixed spectroscopic
pathway decomposition.

This distinction follows directly by differentiating
Eq.~(\ref{eq:LiouvillianEigenproblem}) with respect to a
model parameter $\lambda^\mu$. For nondegenerate
Liouvillian modes, projection onto a left eigenoperator
$\widetilde{\Phi}_m$ with $m\neq n$ gives
\begin{align}
\langle\langle
\widetilde{\Phi}_m
|
\partial_\mu\Phi_n
\rangle\rangle
=
\frac{
\langle\langle
\widetilde{\Phi}_m
|
\partial_\mu\mathcal{L}
|
\Phi_n
\rangle\rangle
}{
\Lambda_n-\Lambda_m
}.
\label{eq:eigenoperatorDerivative}
\end{align}
Thus, the off-diagonal matrix elements of
$\partial_\mu\mathcal{L}$ quantify the local deformation
that mixes distinct Liouvillian eigenoperators. They do
not themselves constitute a geometric connection; rather,
they determine how the local dynamical modes respond to an
infinitesimal change of the physical model.

A parameter-dependent Liouvillian can therefore exhibit
substantial changes in its eigenvalues without changing
its eigenoperator structure. A simple example is the
conventional reduced pure-dephasing model in which the
dephasing rate varies while the dephasing basis remains
fixed. In that case the decay rates change, but no pathway
mixing arises from a rotation of the Liouvillian
eigenoperators. The situation considered here is different:
variation of the pointer-basis orientation changes the
eigenoperator structure itself and thereby produces a
local redistribution among pathway sectors.

Equation~(\ref{eq:eigenoperatorDerivative}) characterizes
this deformation at the level of the instantaneous
Liouvillian modes. For nonlinear spectroscopy, however,
the relevant objects are the finite-time propagators that
connect successive field interactions. We therefore turn
next to the parametric derivative of
$e^{t\mathcal{L}}$, which provides the local response
transport entering the Duhamel construction.

\subsection{Local Liouvillian Deformation and Response Transport}
\label{sec:IIIB}

The local change of the reduced dynamical generator under
an infinitesimal displacement in model space is described
by its parametric derivative. We define the Liouvillian
deformation operator
\begin{align}
\mathcal{X}_\mu
=
\partial_\mu\mathcal{L},
\label{eq:LiouvillianDeformation}
\end{align}
where
$\partial_\mu\equiv\partial/\partial\lambda^\mu$.
The operator $\mathcal{X}_\mu$ specifies how the
Liouvillian changes locally along the coordinate direction
$\lambda^\mu$. Its diagonal matrix elements in the
instantaneous Liouvillian eigenbasis describe changes of
the corresponding eigenvalues, while its off-diagonal
matrix elements generate the eigenoperator mixing
identified in Eq.~(\ref{eq:eigenoperatorDerivative}).
Thus, $\mathcal{X}_\mu$ is a local deformation operator,
rather than a geometric connection. Its relevance to
spectroscopy follows from how this deformation modifies
the finite-time propagators entering the nonlinear
response.

The Liouvillian deformation operator is most naturally
resolved in the instantaneous Liouvillian eigenbasis,
\begin{align}
(\mathcal{X}_\mu)_{mn}
&=
\langle\!\langle
\widetilde{\Phi}_m
|
\mathcal{X}_\mu
|
\Phi_n
\rangle\!\rangle,
\label{eq:DeformationMatrix}
\end{align}
where $\Phi_n$ and $\widetilde{\Phi}_n$ denote the
right and left Liouvillian eigenoperators introduced
in Eq.~(\ref{eq:LiouvillianEigenproblem}). Its diagonal
elements determine the local variation of the
Liouvillian eigenvalues,
\begin{align}
\partial_\mu \Lambda_n
&=
(\mathcal{X}_\mu)_{nn},
\label{eq:EigenvalueDerivative}
\end{align}
whereas its off-diagonal elements determine the local
mixing of distinct eigenoperators through
Eq.~(\ref{eq:eigenoperatorDerivative}). The deformation
matrix therefore separates changes in characteristic
frequencies and decay rates from changes in the
Liouville-space modes through which the spectroscopic
response is resolved.

The distinction between generator deformation and
geometric transport has a familiar analogue in
parameter-dependent Hamiltonian systems. For
\begin{align}
H(\boldsymbol{\lambda})
|n(\boldsymbol{\lambda})\rangle
&=
E_n(\boldsymbol{\lambda})
|n(\boldsymbol{\lambda})\rangle,
\end{align}
the operator $\partial_\mu H$ describes the local
deformation of the Hamiltonian. For nondegenerate
eigenstates, its off-diagonal matrix elements determine
the corresponding change of the eigenvectors,
\begin{align}
\langle m|\partial_\mu n\rangle
&=
\frac{
\langle m|\partial_\mu H|n\rangle
}{
E_n-E_m
},
\qquad m\neq n.
\end{align}
This is directly analogous to
Eq.~(\ref{eq:eigenoperatorDerivative}), with
$\partial_\mu H$ replaced by the Liouvillian deformation
operator $\mathcal{X}_\mu=\partial_\mu\mathcal{L}$.
Neither $\partial_\mu H$ nor $\mathcal{X}_\mu$ is itself
a geometric connection; rather, each determines the
local deformation of the corresponding dynamical
eigenmodes. A geometric connection requires an
additional prescription for comparing local frames at
different points in parameter space, a distinction we
return to below.

The distinction is important. In a Hamiltonian description,
parametric deformation acts on the eigenstates and energies
of a coherent quantum system. In an open-system description,
the corresponding deformation acts on Liouvillian
eigenoperators and eigenvalues and therefore includes
relaxation, decoherence, and environmental pathway mixing
within the same dynamical framework. In the purely unitary
limit,
\begin{align}
\mathcal{L}\rho
&=
-\frac{i}{\hbar}[H,\rho],
\end{align}
the Liouvillian deformation is inherited directly from
the underlying Hamiltonian,
\begin{align}
\partial_\mu\mathcal{L}\,\rho
&=
-\frac{i}{\hbar}
[\partial_\mu H,\rho].
\end{align}
Dissipative contributions extend this construction beyond
the Hamiltonian setting by allowing both the decay rates
and the Liouville-space eigenoperators to vary with the
model parameters. 

In the observational pathway basis, purely Hamiltonian
propagation preserves the pathway decomposition when the
corresponding sectors are invariant under the coherent
generator. Dissipative dynamics that couple these sectors
introduce off-diagonal contributions in the same
decomposition, thereby redistributing amplitude among
pathways that are distinct in the observational basis.
This redistribution constitutes the pathway transport
considered here.

\section{\Rconcept{Minimal Model for Liouvillian Deformation}}
\label{sec:IV}
\subsection{Pointer-Basis Misalignment}
\label{sec:IVA}

The model considered here connects directly to earlier
results for the stationary-state response of open quantum
systems~\cite{Bittner:2026aa,Bittner:2026ab}. In that work,
misalignment between the Hamiltonian eigenbasis and the
environment-selected pointer basis was parameterized by an
angle $\theta$. The resulting stationary-state response
tensor develops an antisymmetric sector associated with
the response geometry of the underlying model manifold.
Here we consider the same physical deformation from a
different perspective. Rather than examining geometric
work accumulated over a quasistatic cycle in parameter
space, we ask how variation of the pointer-basis
orientation modifies a nonlinear spectroscopic response.
The parameter $\theta$ therefore provides a common
deformation coordinate linking the two descriptions:
in the present setting, its variation changes the reduced
Liouvillian and redistributes amplitude among
spectroscopic pathways during the free-evolution
intervals of a multidimensional experiment.

The system consists of a two-level Hamiltonian
\begin{align}
H
&=
\frac{\omega_0}{2}\sigma_z,
\label{eq:H2LS}
\end{align}
whose eigenstates define the observational basis. The
corresponding population and coherence sectors in
Liouville space provide the conventional pathway
decomposition of the nonlinear optical response. For
purely Hamiltonian evolution these sectors are invariant
under the field-free dynamics. Environmental interactions
that are incompatible with this decomposition can instead
couple the sectors and thereby generate the pathway
transport considered below.

\Rone{The reduced dynamics used in all numerical calculations
are generated by
\begin{align}
\mathcal L_\theta[\rho]
={}&
-\frac{i}{\hbar}[H,\rho]
+\gamma_\phi
\left(A_\theta\rho A_\theta-\rho\right)
\nonumber\\
&+
\gamma_1\left(
\sigma_-\rho\sigma_+
-\frac{1}{2}\{\sigma_+\sigma_-,\rho\}
\right),
\label{eq:minimalLindbladian}
\end{align}
where
\begin{align}
A_\theta
&=
\cos\theta\,\sigma_z
+\sin\theta\,\sigma_x
\label{eq:rotatedDephasing}
\end{align}
is the dephasing operator in the rotated pointer basis.
Throughout Figs.~1--3 we use reduced units with
$\hbar=1$ and
\begin{align}
\omega_0=1,\qquad
\gamma_\phi=0.1,\qquad
\gamma_1=0.02.
\end{align}
Thus, $\theta$ controls only the orientation of the
dephasing channel relative to the Hamiltonian eigenbasis,
while the amplitude-relaxation channel proportional to
$\gamma_1$ remains fixed.}

The dephasing channel selects a preferred pointer basis
whose orientation relative to the observational basis is
controlled by $\theta$. The two bases are related by the
unitary transformation
\begin{align}
|p_n(\theta)\rangle
&=
U(\theta)|e_n\rangle,
\label{eq:pointerbasis}
\end{align}
where $|e_n\rangle$ denotes a Hamiltonian eigenstate and
$|p_n(\theta)\rangle$ the corresponding pointer state.
At $\theta=0$ the two bases coincide, whereas finite
$\theta$ introduces a controlled misalignment between
them. Importantly, $\theta$ does not describe a physical
rotation occurring during the dynamics. It is a structural
parameter specifying the orientation of the
environment-selected dephasing basis relative to the
Hamiltonian basis used to define the spectroscopic
pathways.

\Rone{The physical content of $\theta$ is transparent. At
$\theta=0$, the dephasing operator $A_\theta$ is diagonal
in the Hamiltonian basis, so that this channel suppresses
coherences without inducing population transfer in that
basis. The separate amplitude-relaxation channel with rate
$\gamma_1$ remains present and fixed for all values of
$\theta$. As $\theta$ increases, $A_\theta$ becomes
non-diagonal in the Hamiltonian basis and introduces
additional coupling between population and coherence
sectors. Thus $\theta$ does not vary the dissipative rates;
rather, it controls the basis incompatibility of the
dephasing channel and the resulting pathway mixing.}

The basis mismatch becomes dynamically relevant through
the Liouvillian governing the open-system evolution,
\begin{align}
\dot{\rho}
&=
\mathcal{L}(\bm{\lambda})\rho,
\end{align}
where $\bm{\lambda}$ denotes the model-space parameters
that characterize the reduced dynamics.
The spectral reconstruction in Sec.~\ref{sec:V} follows a
one-dimensional slice, varying $\theta$ at fixed
$\gamma_\phi=0.1$. To examine the model geometry, we also
regard the dephasing rate as an independent control and use
the two coordinates
\begin{align}
q^1&=\theta,\qquad q^2=\gamma_\phi/0.1,
\end{align}
over $0\leq\theta\leq\pi/2$ and
$0.05\leq\gamma_\phi\leq0.20$, with $\gamma_1=0.02$ and
$\omega_0=1$. The full-domain convergence study in Sec.~S2.2 of the
Supporting Information uses the same two-control domain.

\begin{figure*}
\centering
\includegraphics[width=0.94\textwidth]{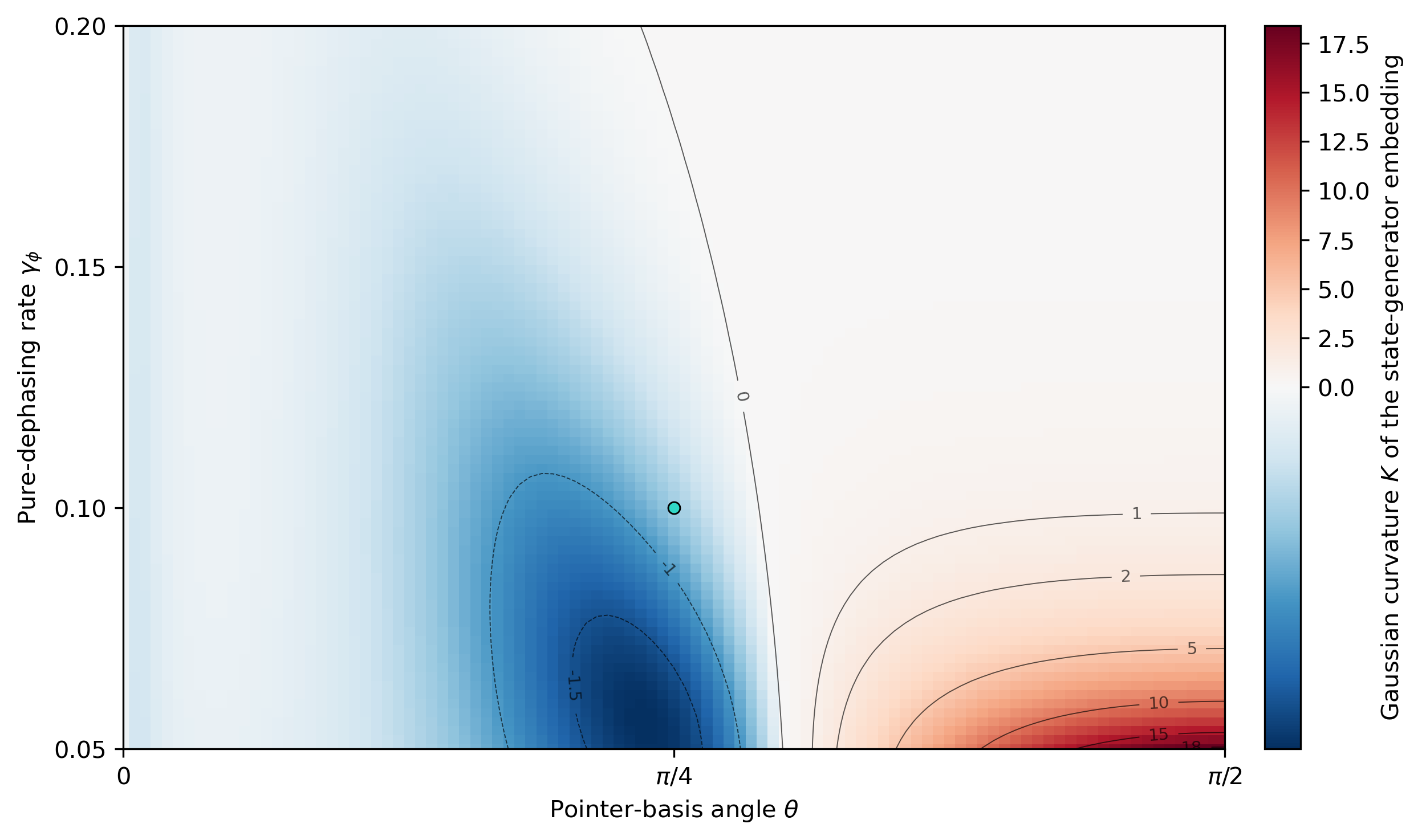}
\caption{Gaussian curvature of the two-control state--generator
manifold over $(\theta,\gamma_\phi)$. The embedding concatenates
the real and imaginary components of the normalized stationary
state and the Liouvillian, with $\omega_0=1$ and fixed
$\gamma_1=0.02$. The marker identifies $(\theta,\gamma_\phi)
=(\pi/4,0.1)$. Curvature is evaluated on a $101\times81$ grid
using central differences with step $10^{-4}$ in coordinates
$(\theta,\gamma_\phi/0.1)$.}
\label{fig:liouvillian-manifold}
\end{figure*}

For the geometric visualization, we embed each model using
its normalized stationary state and its Liouvillian,
\begin{align}
\Phi(q)=\big(\operatorname{Re}\rho_{\rm ss},
\operatorname{Im}\rho_{\rm ss},
\operatorname{Re}\mathcal L,
\operatorname{Im}\mathcal L\big),\\
g_{\mu\nu}(q)=\partial_\mu\Phi\cdot\partial_\nu\Phi,
\label{eq:state-generator-embedding}
\end{align}
where the matrix components are vectorized and concatenated
as real coordinates. The Gaussian curvature $K$ is computed
from this induced metric. Figure~\ref{fig:liouvillian-manifold}
shows that the surface is nonuniformly curved, with the
largest sampled curvature at large $\theta$ and low
$\gamma_\phi$.

Expanding locally about the aligned reference point
$\bm{\lambda}=0$,
\begin{align}
\mathcal{L}(\bm{\lambda})
&=
\mathcal{L}_0
+
\sum_\mu \lambda^\mu \mathcal{X}_\mu
+
\frac{1}{2}\sum_{\mu\nu}
\lambda^\mu\lambda^\nu \mathcal{Y}_{\mu\nu}
+
O(\lambda^3),
\end{align}
where
\begin{align}
\mathcal{X}_\mu
&=
\left.
\frac{\partial\mathcal{L}}
{\partial\lambda^\mu}
\right|_{\bm{\lambda}=0},
\qquad
\mathcal{Y}_{\mu\nu}
=
\left.
\frac{\partial^2\mathcal{L}}
{\partial\lambda^\mu\partial\lambda^\nu}
\right|_{\bm{\lambda}=0}.
\end{align}
The operators $\mathcal{X}_\mu$ describe the leading
local deformation of the Liouvillian away from the
reference dynamics, while $\mathcal{Y}_{\mu\nu}$
describe the corresponding second-order corrections.
Their effect on the finite-time propagators, and hence
on the spectroscopic pathways, is obtained through the
Duhamel expansion developed below. For the minimal
one-parameter model, the expansion reduces to
\begin{align}
\mathcal{L}(\theta)
&=
\mathcal{L}_0
+
\theta\mathcal{X}
+
\frac{\theta^2}{2}\mathcal{Y}
+
O(\theta^3).
\label{eq:ThetaExpansion}
\end{align}

\Rone{For the rotated pure-dephasing model,
$\theta$ enters only through the dissipative operator,
so that
\begin{align}
\mathcal X
=
\left.
\frac{\partial\mathcal L(\theta)}
{\partial\theta}
\right|_{\theta=0}
=
\left.
\frac{\partial\mathcal D_\theta}
{\partial\theta}
\right|_{\theta=0}.
\label{eq:XexplicitMatrix}
\end{align}
Thus $\mathcal X$ has a direct physical interpretation:
it is the leading change in the reduced generator produced
by an infinitesimal rotation of the bath-selected pointer
basis away from the Hamiltonian basis. In the
observational Liouville basis
$\{|e_0e_0\rangle\!\rangle,
  |e_0e_1\rangle\!\rangle,
  |e_1e_0\rangle\!\rangle,
  |e_1e_1\rangle\!\rangle\}$,
its off-diagonal matrix elements explicitly couple the
population and coherence sectors.}

\Rone{For example, in the column-stacked Liouville basis
\begin{align}
\operatorname{vec}\rho
=
(\rho_{gg},\rho_{eg},\rho_{ge},\rho_{ee})^T,
\end{align}
the first-order deformation operator for the present model is
\begin{align}
\mathcal X
&=
\left.
\frac{\partial\mathcal L_\theta}{\partial\theta}
\right|_{\theta=0}
\nonumber\\
&=
\gamma_\phi
\begin{pmatrix}
0&-1&-1&0\\
-1&0&0&1\\
-1&0&0&1\\
0&1&1&0
\end{pmatrix}.
\label{eq:Xexplicit}
\end{align}
Its off-diagonal structure makes explicit that an
infinitesimal rotation of the dephasing basis couples
population and coherence sectors of the reference
Liouvillian.}

\Rone{Because the amplitude-relaxation term is independent
of $\theta$, it contributes to the reference propagator
$\mathcal L_0$ but not to $\mathcal X$.  The local
$\theta$-derivative therefore isolates the pathway mixing
generated specifically by rotation of the dephasing basis.}

We begin by defining the third-order rephasing response
of the aligned reference model,
\begin{align}
S_{\mathrm{R}}^{(0)}
(\omega_3,t_2,\omega_1)
&=
\sum_{a=1}^{2}
S_{a,\mathrm{R}}^{(0)}
(\omega_3,t_2,\omega_1),
\label{eq:S0Decomposition}
\end{align}
\Rone{where the superscript $(0)$ denotes evaluation at
the aligned reference point $\theta=0$. For the present
two-level system, application of the rotating-wave
approximation, phase-matching condition, and terminal
trace leaves four double-sided Feynman diagrams
contributing to the third-order response
\cite{Mukamel1995}. Two occur in
the rephasing channel,
\begin{align}
\mathbf{k}_{\rm R}
&=
-\mathbf{k}_1+\mathbf{k}_2+\mathbf{k}_3,
\end{align}
and two in the nonrephasing channel,
\begin{align}
\mathbf{k}_{\rm NR}
&=
\mathbf{k}_1-\mathbf{k}_2+\mathbf{k}_3.
\end{align}
Within each channel, the two surviving diagrams
correspond to ground-state bleach and stimulated
emission. Equation~(\ref{eq:S0Decomposition}) contains
the two diagrams belonging to the rephasing channel
considered here. Each contribution is Fourier
transformed with respect to the first and third
coherence intervals, while $t_2$ is the waiting time
between the second and third field interactions. The
reference pathways are propagated with the complete
open-system Liouvillian $\mathcal L_0$ and therefore
include both dephasing and the fixed
amplitude-relaxation channel.}

\Rone{Finite pointer-basis misalignment deforms the
spectroscopic response by introducing additional
population--coherence mixing through the rotated
dephasing channel. Rather than constructing an explicit
finite pathway-transport operator, we evaluate this
deformation locally by expanding the Liouvillian-dependent
propagators about the aligned reference point
$\theta=0$. The resulting propagator derivatives are then
inserted directly into the nonlinear response sequence
and collected order by order in $\theta$. Thus, the
spectroscopic response is reconstructed as
\begin{align}
S_{\mathrm R}(\theta)
&=
S_{\mathrm R}^{(0)}
+\theta S_{\mathrm R}^{(1)}
+\frac{\theta^2}{2}S_{\mathrm R}^{(2)}
+O(\theta^3),
\label{eq:ResponseExpansion}
\end{align}
where $S_{\mathrm R}^{(1)}$ and
$S_{\mathrm R}^{(2)}$ contain all first- and
second-order changes of the Liouvillian propagators
appearing in the response. These terms describe the
spectroscopic consequences of the pathway mixing induced
by the local Liouvillian deformation.}

To construct the local response expansion, we first
expand the finite-time Liouvillian propagator
\begin{align}
U_\theta(\tau)
&=
e^{\mathcal{L}(\theta)\tau}
\end{align}
about the aligned reference point $\theta=0$. The
Duhamel identity~\cite{Breuer2002,Wilcox1967,Hall2015}
expresses the derivatives of the propagator in terms of
the reference dynamics and the local Liouvillian
deformations $\mathcal{X}$ and $\mathcal{Y}$ defined in
Eq.~(\ref{eq:ThetaExpansion}). Writing
\begin{align}
U_\theta(\tau)
&=
U_0(\tau)
+\theta U_1(\tau)
+\frac{\theta^2}{2}U_2(\tau)
+O(\theta^3),
\label{eq:GeneralDuhamel}
\end{align}
with
\begin{align}
U_0(\tau)
&=
e^{\mathcal{L}_0\tau},
\end{align}
the first derivative is
\begin{align}
U_1(\tau)
&=
\int_0^\tau ds\,
e^{\mathcal{L}_0(\tau-s)}
\mathcal{X}
e^{\mathcal{L}_0s}.
\label{eq:FirstOrderGeneral}
\end{align}
Thus, the first-order correction corresponds to a
single insertion of the local Liouvillian deformation
$\mathcal{X}$ within the reference evolution. The
second derivative is
\begin{align}
U_2(\tau)
&=
\int_0^\tau ds\,
e^{\mathcal{L}_0(\tau-s)}
\mathcal{Y}
e^{\mathcal{L}_0s}
\nonumber\\
&\quad+
2\int_0^\tau ds
\int_0^s ds'\,
e^{\mathcal{L}_0(\tau-s)}
\mathcal{X}
e^{\mathcal{L}_0(s-s')}
\mathcal{X}
e^{\mathcal{L}_0s'}.
\label{eq:SecondOrderGeneral}
\end{align}
The two contributions to $U_2$ have distinct origins:
the first contains a single insertion of the
second-order Liouvillian deformation $\mathcal{Y}$,
whereas the second contains two time-ordered insertions
of the first-order deformation $\mathcal{X}$. Together,
they give the complete second derivative of the
finite-time propagator with respect to $\theta$,
evaluated at $\theta=0$.

The finite-time propagator governs evolution during the
waiting interval $t_2$. The first and third coherence
intervals are treated in the frequency domain, where the
corresponding Liouvillian resolvents admit an analogous
expansion about $\theta=0$.

Defining the frequency-domain propagator as
\begin{align}
G_\theta(\omega)
&=
-\left[
\mathcal{L}(\theta)+i\omega-\eta
\right]^{-1},
\end{align}
we write
\begin{align}
G_\theta(\omega)
&=
G_0(\omega)
+\theta G_1(\omega)
+\frac{\theta^2}{2}G_2(\omega)
+O(\theta^3),
\end{align}

Defining the frequency-domain propagator as
\begin{align}
G_\theta(\omega)
&=
-\left[
\mathcal{L}(\theta)+i\omega-\eta
\right]^{-1},
\label{eq:ResolventDefinition}
\end{align}
where $\eta>0$ provides the usual convergence
broadening, we expand
\begin{align}
G_\theta(\omega)
&=
G_0(\omega)
+\theta G_1(\omega)
+\frac{\theta^2}{2}G_2(\omega)
+O(\theta^3),
\label{eq:ResolventExpansion}
\end{align}
with
\begin{align}
G_0(\omega)
&=
-\left[
\mathcal{L}_0+i\omega-\eta
\right]^{-1},
\label{eq:G0}
\\
G_1(\omega)
&=
G_0(\omega)\mathcal{X}G_0(\omega),
\label{eq:G1}
\\
G_2(\omega)
&=
2G_0(\omega)\mathcal{X}G_0(\omega)
\mathcal{X}G_0(\omega)
+
G_0(\omega)\mathcal{Y}G_0(\omega).
\label{eq:G2}
\end{align}
Thus, as for the finite-time propagator, the
second-order resolvent correction contains both a
double insertion of the first-order Liouvillian
deformation $\mathcal{X}$ and a single insertion of
the second-order deformation $\mathcal{Y}$.


The Duhamel expansion plays the role of a local
perturbation theory for the spectroscopic response.
The deformation operators $\mathcal{X}$ and
$\mathcal{Y}$ characterize the first- and second-order
variation of the Liouvillian about the aligned reference
point, while the corresponding derivatives of the
finite-time propagator and frequency-domain resolvents
determine the resulting changes in the nonlinear
response. In this way, the spectral deformation is
related directly to the local parametric dependence of
the open-system generator rather than introduced through
phenomenological corrections. The following section
examines how accurately this second-order local
reconstruction captures the nonlinear spectrum as the
pointer-basis misalignment is increased away from
$\theta=0$.

\section{\Rconcept{Spectroscopic Signatures of Parametric Response Transport}}
\label{sec:V}

\Rconcept{The local differential framework developed above shows how
Liouvillian deformations generate transport between
observational pathways. This transport modifies the
nonlinear spectroscopic response. The calculations below
therefore quantify local response deformation and pathway
transport in the minimal two-level model introduced in
Sec.~\ref{sec:IVA}. The convergence analysis then tests
successive response orders in this two-state model and in
a three-state bright-state model. Unless stated otherwise,
errors are relative integrated $L^2$ errors of the
absorptive response; the error definition and spectral
conventions are given in Sec.~S1 of the Supporting
Information.}
\Rone{Although the underlying observable is the full
two-dimensional rephasing response
$S_{\mathrm R}(\omega_3,t_2,\omega_1)$, we focus below
on its diagonal restriction,
\begin{align}
S_{\mathrm R}^{\mathrm{diag}}(\omega;t_2)
\equiv
S_{\mathrm R}(\omega,t_2,\omega).
\label{eq:DiagonalResponse}
\end{align}
This restriction is made for visualization rather than
as an additional approximation to the response
calculation.  Displaying the diagonal response makes it
possible to follow directly how the absorptive and
dispersive spectral features evolve as the Liouvillian
is transported away from the aligned reference model.
The full two-dimensional response is calculated before
the diagonal slice is selected.}

\Rone{For all results shown below, the waiting time is
fixed at $t_2=1$ in reduced units with $\omega_{0} = 1$.  The frequency-domain
response is evaluated over
$\omega_1,\omega_3\in[0.25,1.75]$, with resolvent
regularization $\eta=10^{-3}$. Thus the curves displayed
below correspond to
$S_{\mathrm R}(\omega,t_{2}=1,\omega;\theta)$, selected from
the full two-dimensional rephasing response.}

In practice, the mismatch between the observational pathway basis and the
effective reduced Liouvillian is not observed directly but is inferred
from systematic deviations between the measured spectrum and that
predicted by a reference Liouvillian. Figure~\ref{fig:TransportError} provides a concrete
illustration of this inverse problem. Beginning with a reference
Liouvillian, the Duhamel transport expansion reconstructs the spectral
distortions produced by an unknown basis mismatch. \Rconcept{The
reconstructed response derivatives therefore provide a quantitative
measure of the local pathway transport generated by the basis
mismatch. A geometric connection for comparing such local information
between distinct models requires the additional state--generator
structure described in Ref.~\citenum{Bittner:2026ab}.}

\subsection{Diagonal Spectral Response}
\label{sec:VA}

\begin{figure*}
\centering
\begin{minipage}{0.48\textwidth}
\centering
\includegraphics[width=\linewidth]{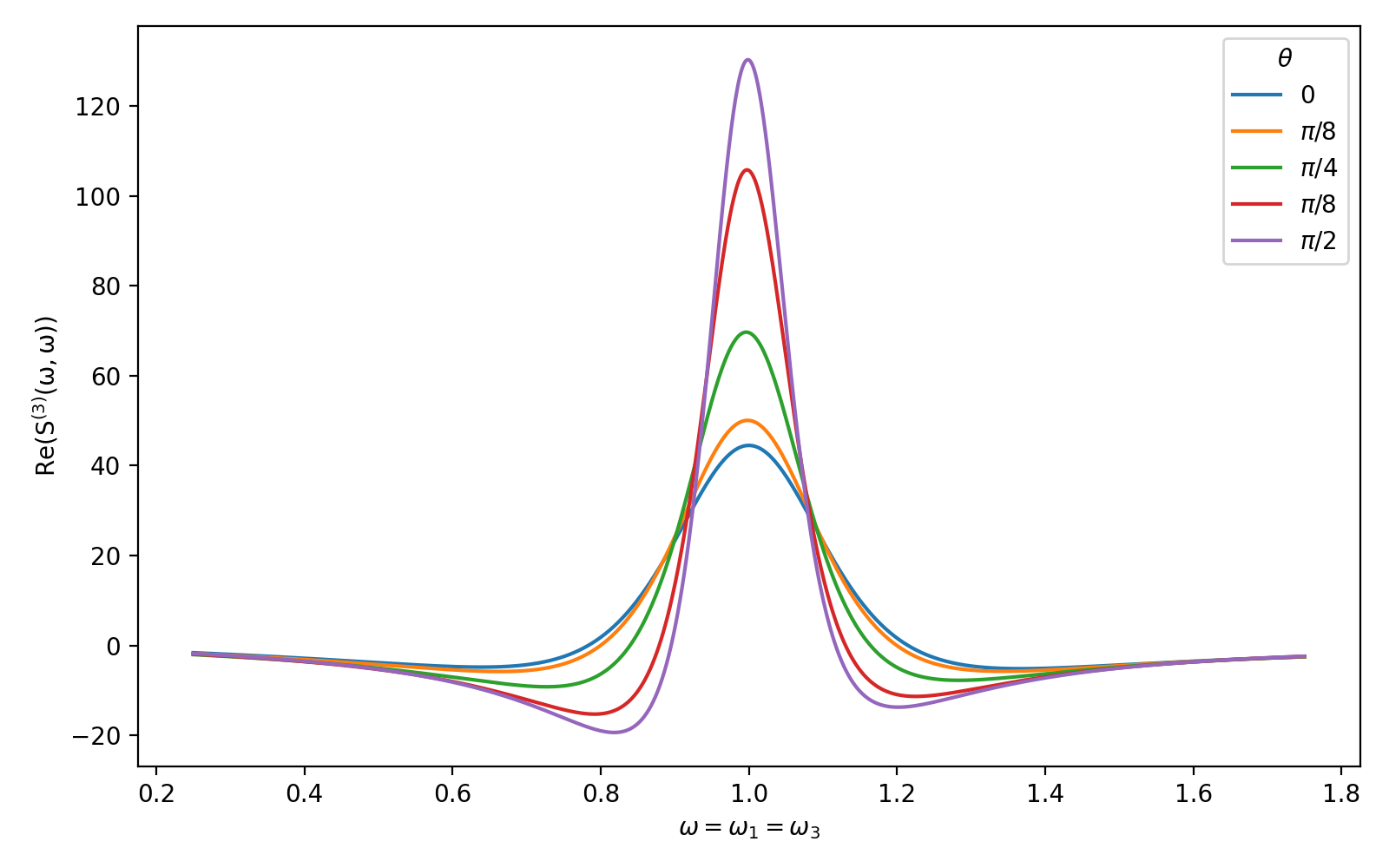}
\end{minipage}
\hfill
\begin{minipage}{0.48\textwidth}
\centering
\includegraphics[width=\linewidth]{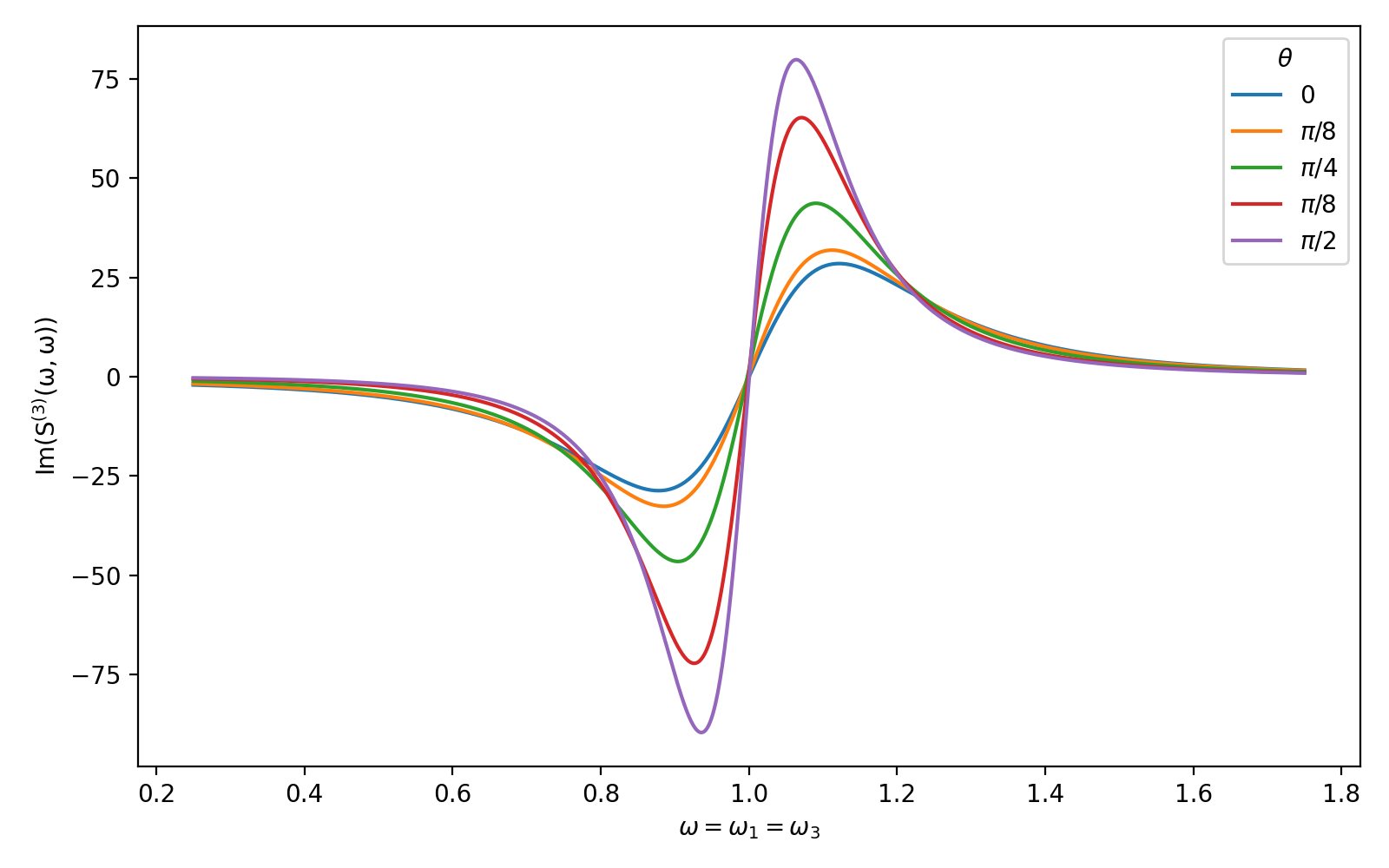}
\end{minipage}
\caption{Real (left) and imaginary (right) components 
of the diagonal third-order rephasing response
$S_{\mathrm R}(\omega,t_{2}=1,\omega;\theta)$ for
$\theta=0,\pi/8,\pi/4,3\pi/8,$ and $\pi/2$.
\Rone{At $\theta=0$, the dephasing and Hamiltonian
bases coincide, so the dephasing channel does not mix
population and coherence sectors; the fixed
amplitude-relaxation channel with $\gamma_1=0.02$
remains present. Negative features in the real response
at $\theta=0$ are dispersive components of the causal
complex response, related to the corresponding absorptive
structure through the Kramers--Kronig relations, and do
not indicate additional population transfer induced by
basis misalignment.}
For $\theta\neq0$, rotation of the dephasing basis
introduces additional population--coherence mixing and
produces continuous, structured deformation of the
complex spectrum.}
\label{fig:DiagResponse}
\end{figure*}

Figure~\ref{fig:DiagResponse} shows the real and 
imaginary components of the diagonal response for 
increasing pointer-basis misalignment. The Hamiltonian, 
optical transition operators, dephasing rate, and
amplitude-relaxation rate are held fixed throughout.
\Rone{The only parameter varied is the orientation
$\theta$ of the dephasing operator relative to the
Hamiltonian basis. Consequently, the spectral evolution
with $\theta$ isolates the effect of this basis
misalignment against the fixed open-system reference
dynamics.}

As the pointer basis departs from the observational 
basis, both the real and imaginary components of the 
response evolve continuously and in a highly 
structured manner. The real component tracks the 
redistribution of spectral weight among the 
contributing pathways, while the imaginary component 
is particularly sensitive to the accompanying phase 
relationships. Basis misalignment does more than 
broaden or attenuate the response --- it transports 
pathway weight through Liouville space, producing 
measurable deformations of the complex spectrum. 
\Rconcept{These deformations are direct spectroscopic signatures of
parametric Liouvillian deformation and pathway transport.}

\subsection{Accuracy of the Transport Expansion}
\label{sec:VB}

\begin{figure}
\centering
\includegraphics[width=0.75\linewidth]{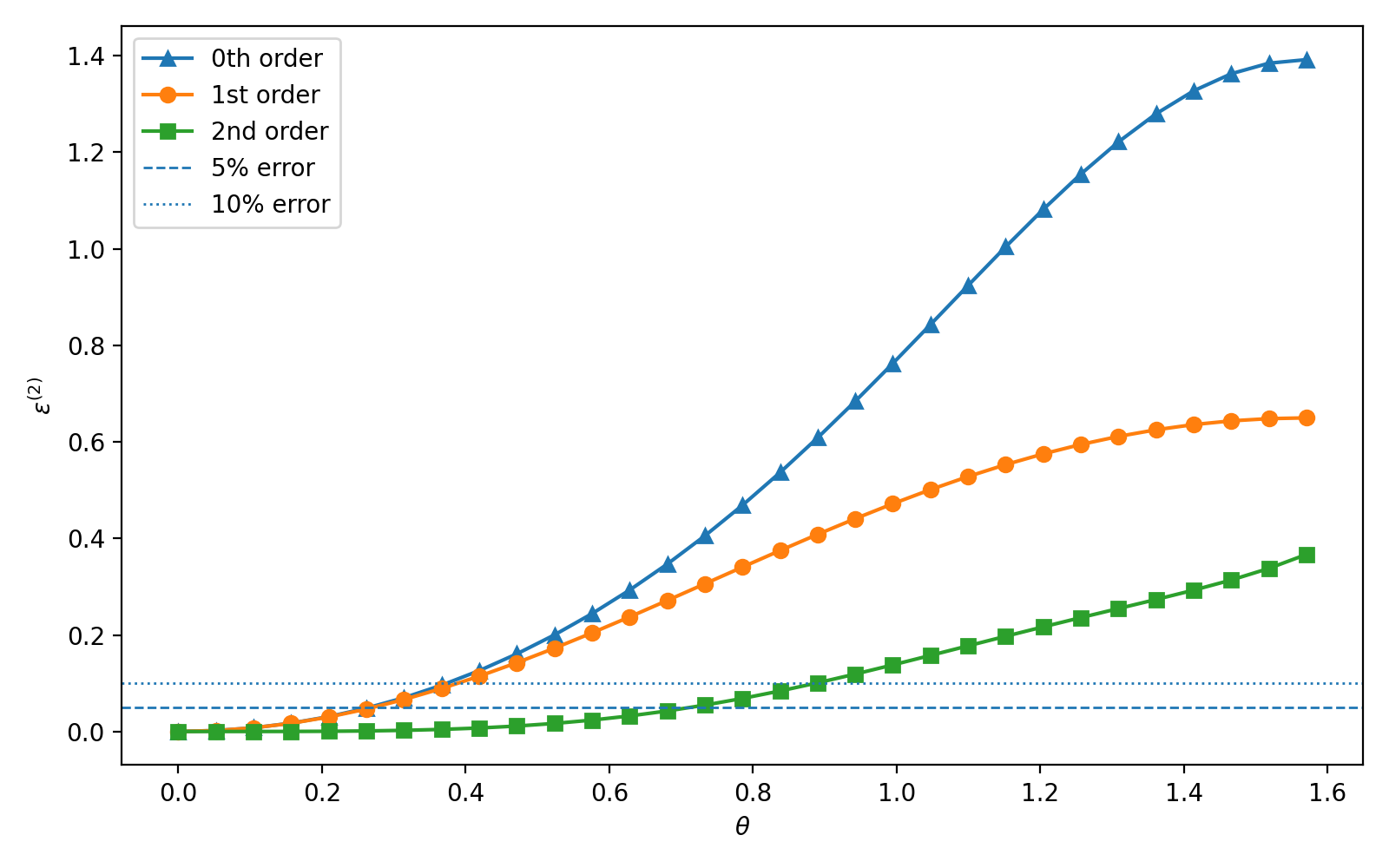}
\caption{Global reconstruction error $\epsilon^{(n)}$
for the zeroth-, first-, and second-order Duhamel
expansions as a function of pointer-basis misalignment
$\theta$. The expansion is constructed locally about
the aligned reference model $\theta=0$. The second-order
result substantially reduces the global error over the
entire interval and retains a useful reconstruction even
at the largest displacement, $\theta=\pi/2$, where the
global error is approximately $37\%$. The progressive
increase in error with $\theta$ reflects the expected
limitation of extrapolating a finite-order local expansion
far from its reference point.}
\label{fig:TransportError}
\end{figure}

\Rone{Figure~\ref{fig:TransportError} quantifies how far
the local Duhamel expansion can be extrapolated from its
reference point.  The first-order approximation captures
the leading deformation of the response, while inclusion
of the second-order terms substantially reduces the
global reconstruction error throughout the interval.
As expected for a local expansion about $\theta=0$, the
error grows as the target Liouvillian is moved farther
from the reference model. At maximal basis misalignment,
$\theta=\pi/2$, the second-order global error has grown
to approximately $37\%$.}

\Rone{Perhaps more significant than this eventual growth
in error is the range over which such a low-order local
expansion remains informative. The second-order
construction uses only derivatives evaluated at
$\theta=0$, yet continues to retain the dominant spectral
structure as far as $\theta=\pi/2$. Thus the calculation
should not be interpreted as demonstrating uniform
pointwise accuracy over the full parameter interval.
Rather, it shows that local differential information
about the Liouvillian contains sufficient structure to
reconstruct the principal spectral deformation over a
surprisingly large displacement in model space.}

\Rconcept{This robustness is the key quantitative result of
this section. Although the Duhamel expansion is
derived from local Liouvillian deformations about
$\theta = 0$, it remains informative well beyond the
regime where a weak-deformation approximation might
be expected to hold. The objective is not merely
pointwise spectral reconstruction, but identification of
the local deformation of the reduced Liouvillian
responsible for the observed spectrum. The dominant
spectroscopic consequences of pathway transport are
therefore captured by a compact set of response-transport
coefficients that quantify the underlying transport
mechanism.}

\Rone{The calculation therefore provides a stringent test
of the local response construction. The Duhamel
coefficients encode how the nonlinear spectrum responds
to infinitesimal deformation of the reference
Liouvillian, while Fig.~\ref{fig:TransportError} tests
how far that local information can be extrapolated.
The eventual deterioration at large $\theta$ identifies
the intrinsic limitation of reconstruction from a single
reference point and motivates replacing long-range
extrapolation by a sequence of local transports through
parameter space.}

\begin{figure*}
\centering
\includegraphics[width=0.48\linewidth]{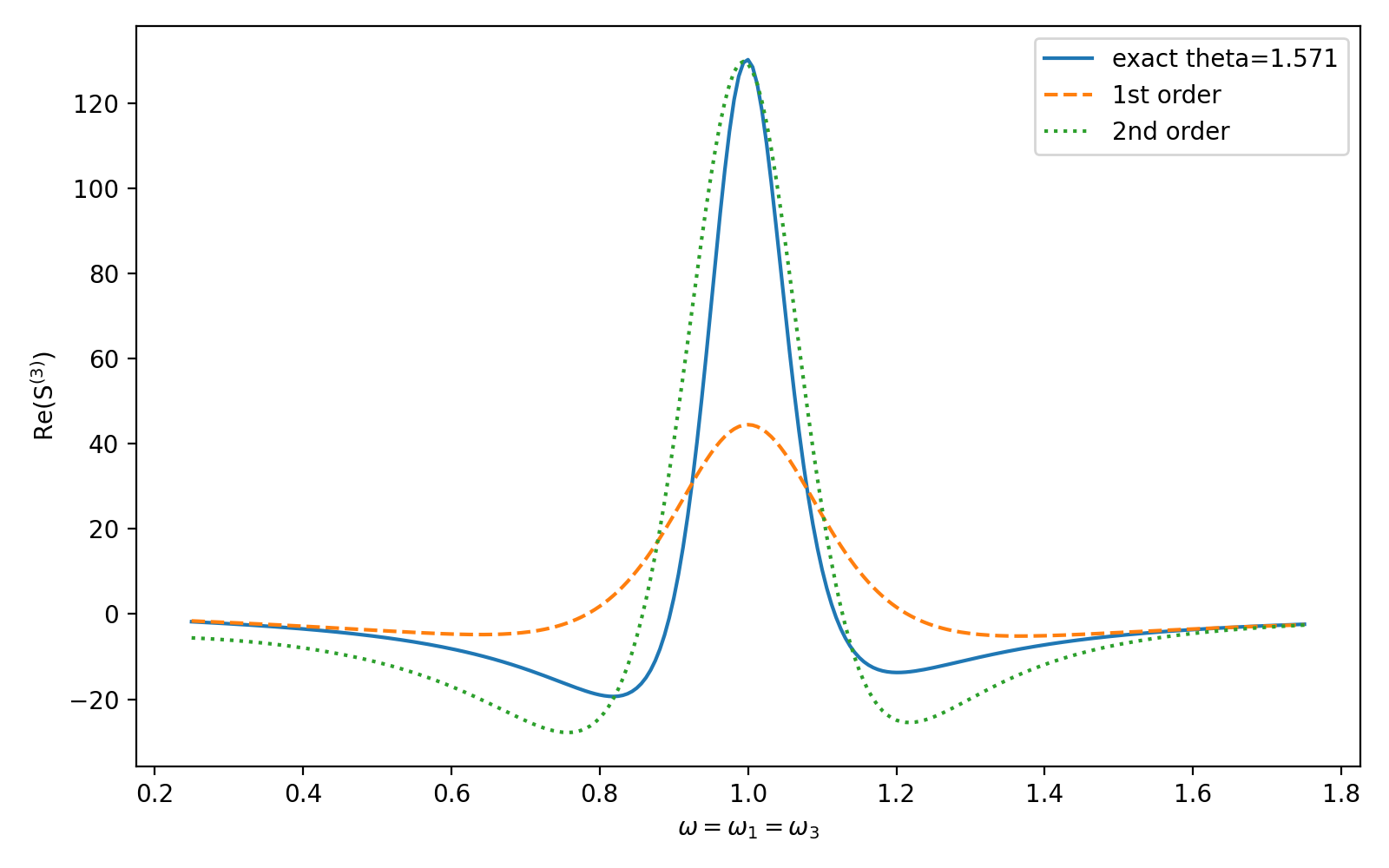}
\includegraphics[width=0.48\linewidth]{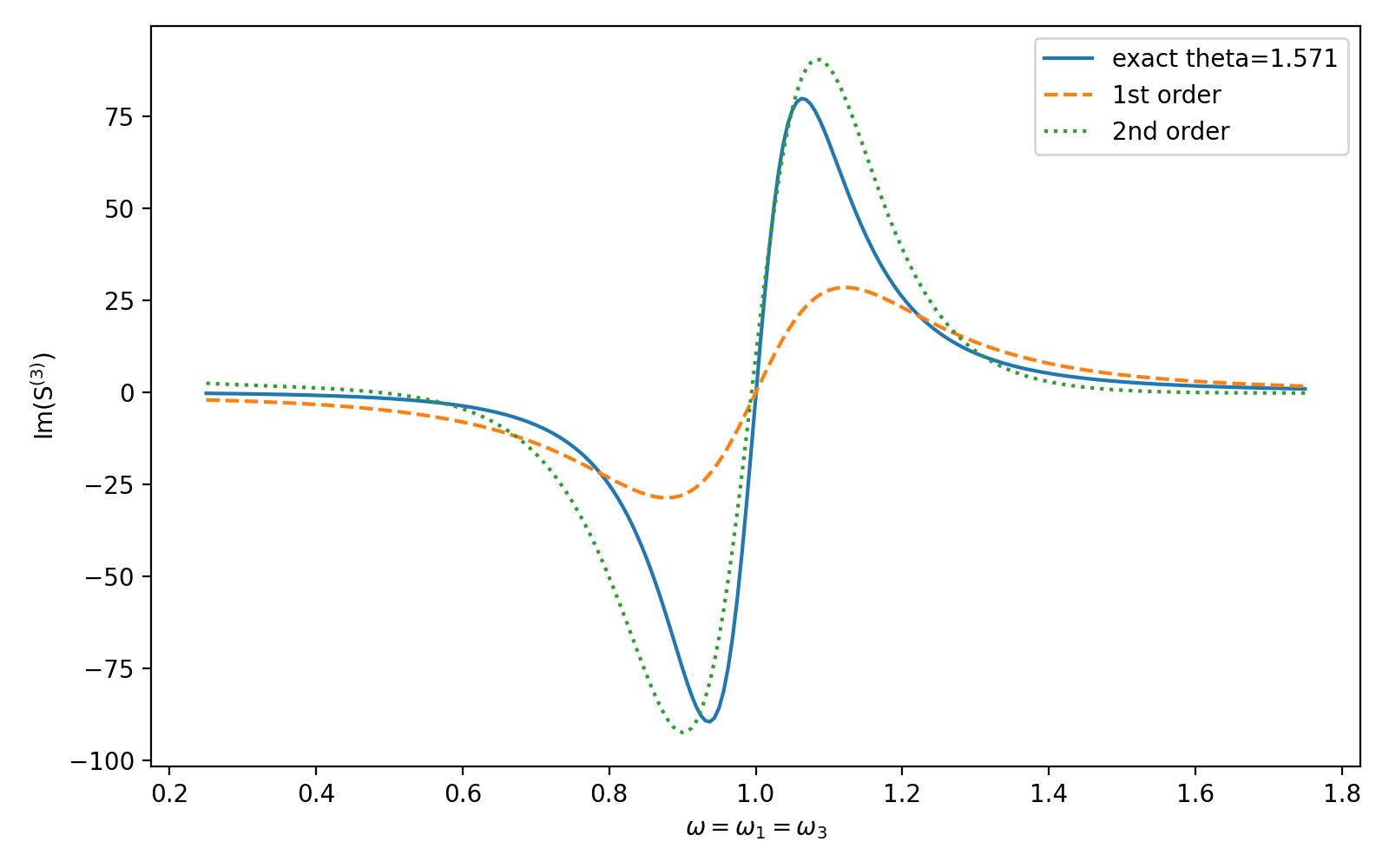}
\caption{Comparison between the exact and reconstructed
diagonal responses at maximal misalignment,
$\theta=\pi/2$. The second-order Duhamel reconstruction
retains the dominant resonant and dispersive structure
of the exact response despite being constructed entirely
from local derivatives at $\theta=0$. Quantitative
pointwise deviations remain, consistent with the
approximately $37\%$ global reconstruction error at
this largest parameter displacement.}
\label{fig:worst-case}
\end{figure*}

\Rone{The most demanding reconstruction is shown in
Fig.~\ref{fig:worst-case}, where the target model lies
a distance $\pi/2$ from the point about which the
Duhamel expansion is constructed. The first-order
approximation captures the overall resonant structure
but substantially underestimates its amplitude and does
not reproduce the detailed line shape. Inclusion of the
second-order terms recovers the dominant features of
both the real and imaginary responses, including the
central resonance and dispersive structure, although
significant local deviations remain. These deviations
are genuine higher-order truncation effects expected
when a second-order local expansion is extrapolated over
such a large parameter displacement.}

\Rone{Large relative deviations at isolated frequencies
should also be interpreted with care when the exact
response passes through a small value, since normalization
by a small local signal can amplify the corresponding
percentage error.  The global error in
Fig.~\ref{fig:TransportError} and the direct line-shape
comparison in Fig.~\ref{fig:worst-case} therefore provide
complementary measures of reconstruction quality.}

\subsection{Order-by-order convergence in the two-state model}
\label{sec:two-state-convergence}

We next assess the range over which successive response orders improve the
reconstruction in the two-state model. The response was expanded about the
midpoint $(\theta_0,\gamma_{\phi,0})=(\pi/4,0.125)$, with the exact response
at the reference point fixed as the constant term. First-, second-, and
third-order coefficients were obtained from least-squares fits to exact maps
on a symmetric local stencil; they are numerical response derivatives rather
than analytic Fr\'echet derivatives. We compare the resulting absorptive maps
with direct calculations over the full rectangle
$0\leq\theta\leq\pi/2$ and $0.05\leq\gamma_\phi\leq0.20$, sampled on a
$21\times21$ parameter grid. Table~\ref{tab:two-state-convergence} reports
the mean relative integrated $L^2$ error over that grid. The fitted-stencil
details and full error maps, including the mean and worst-case errors, are
reported in Sec.~S2.2 of the Supporting Information.

\begin{table}
\caption{Mean relative integrated $L^2$ error (percent) of the two-state
absorptive response over the full $21\times21$ parameter grid, for successive
orders of the expansion about $(\theta_0,\gamma_{\phi,0})=(\pi/4,0.125)$.}
\label{tab:two-state-convergence}
\begin{ruledtabular}
\begin{tabular}{rccc}
$t_2$ & First order & Second order & Third order \\
\hline
0.2 & 26.3 & 14.9 & 10.2 \\
1   & 24.3 & 13.9 & 9.2 \\
5   & 34.5 & 19.0 & 14.3 \\
\end{tabular}
\end{ruledtabular}
\end{table}

\begin{figure*}
\centering
\includegraphics[width=0.96\textwidth]{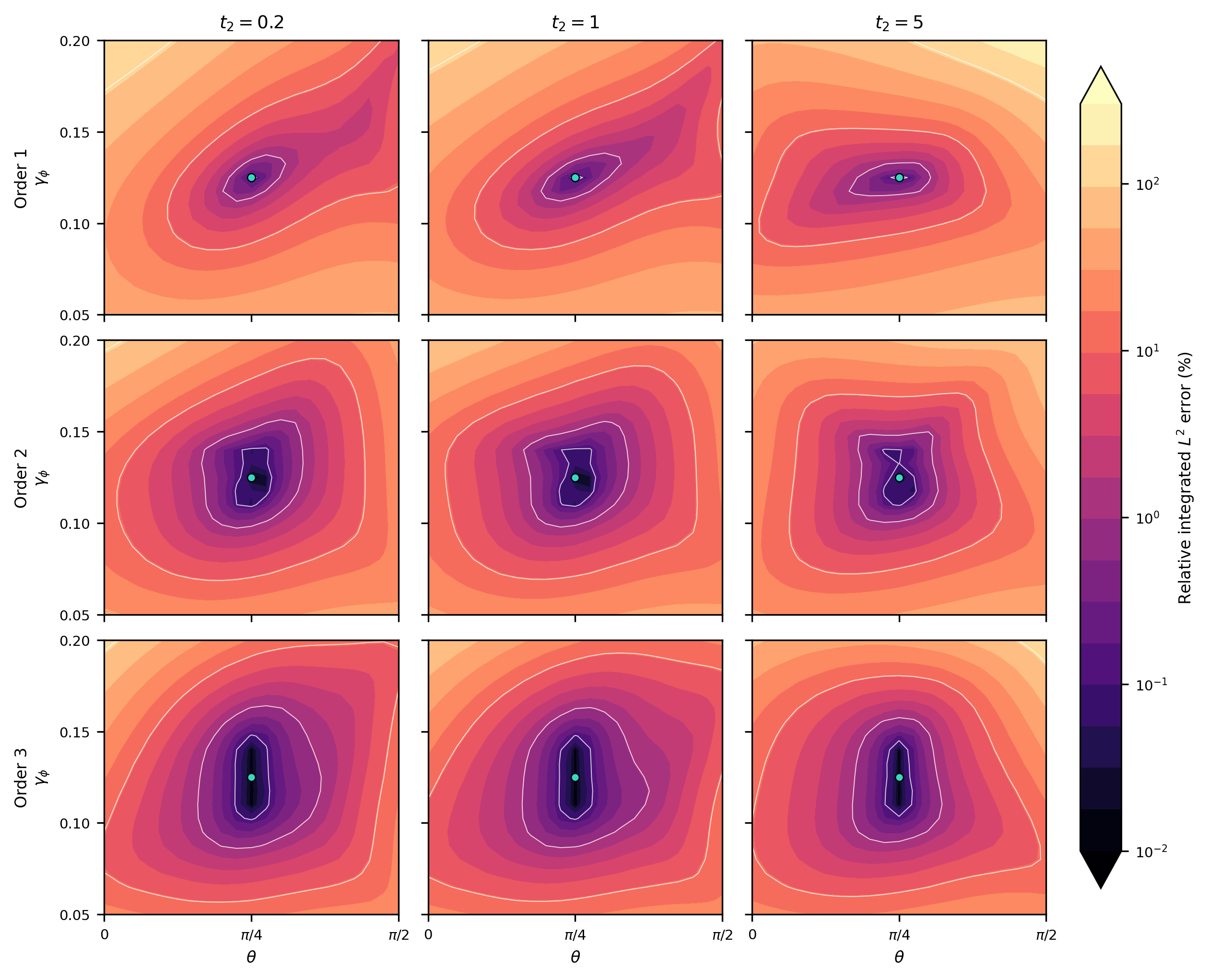}
\caption{Relative integrated $L^2$ error (percent) of the two-state
absorptive response across the $(\theta,\gamma_\phi)$ domain. Rows show
first-, second-, and third-order expansions about
$(\pi/4,0.125)$; columns show $t_2=0.2$, 1, and 5. The marker identifies the
expansion reference. The filled contours use a logarithmic color scale.}
\label{fig:two-state-convergence}
\end{figure*}

Figure~\ref{fig:two-state-convergence} resolves this order-by-order trend
across the two-dimensional parameter domain.

The domain-averaged error decreases at each order for all three waiting
times. This trend is not uniform across parameter space: the maximum error
over the grid at third order is still 96.6\%--144.4\%, and at $t_2=5$ it
exceeds the second-order maximum. By contrast, in the smaller neighborhood
$0.589\leq\theta\leq0.982$ and
$0.1125\leq\gamma_\phi\leq0.1375$, the second-order error remains below
0.96\% at every sampled point for all three waiting times. Refining that
neighborhood from a $5\times5$ to a $21\times21$ parameter grid leaves the
maximum errors unchanged to the reported precision. Thus the two-state
expansion converges rapidly near its reference model, while higher order
improves the average full-domain reconstruction without ensuring accurate
extrapolation at the parameter-space edges. The finite-difference steps and
grid-refinement comparison for the local neighborhood are given in
Sec.~S2.3 of the Supporting Information.

\subsection{Local convergence in the three-state model}
\label{sec:three-state-convergence}

We tested the same order-by-order behavior in a three-state bright-state
model with states $|G\rangle$, $|B_1\rangle$, and $|B_2\rangle$. The
deformation parameter $\lambda=\kappa$ is the one-way internal-conversion
rate from $|B_2\rangle$ to $|B_1\rangle$. Here the response coefficients are
computed analytically by Duhamel insertions, including Neumann expansions of
the coherence-interval resolvents and time-ordered insertions in the waiting
interval. We evaluate the expansion about $\lambda_{\mathrm{ref}}=0.054$
at the nearby exact target $\lambda=0.060$, so that
$\Delta\lambda=0.006$. Table~\ref{tab:three-state-convergence} gives the
relative integrated $L^2$ errors for the absorptive maps. The model
parameters and analytic resolvent and waiting-time coefficients are
described in Sec.~S3.1 of the Supporting Information, with the local
order-by-order results detailed in Sec.~S3.2.

\begin{table}
\caption{Relative integrated $L^2$ error (percent) of the three-state
absorptive response at $\lambda=0.060$, reconstructed from a local Duhamel
expansion about $\lambda_{\mathrm{ref}}=0.054$.}
\label{tab:three-state-convergence}
\begin{ruledtabular}
\begin{tabular}{rccc}
$t_2$ & First order & Second order & Third order \\
\hline
0.2 & 0.770 & 0.0670 & 0.00574 \\
1   & 0.795 & 0.0694 & 0.00595 \\
5   & 0.932 & 0.0825 & 0.00714 \\
10  & 1.093 & 0.0990 & 0.00868 \\
\end{tabular}
\end{ruledtabular}
\end{table}

For this short displacement, each added order reduces the error by roughly
an order of magnitude, and the third-order map agrees with the exact target
to better than $0.009\%$ at every waiting time. The result depends on using
a nearby reference. Expanding instead from zero transfer to the same
$\lambda=0.060$ target does not give useful order-by-order convergence: at
$t_2=0.2$, the first-, second-, and third-order errors are 991\%, 3238\%,
and 10\,274\%, respectively, and at $t_2=10$ they rise to 1550\%, 5238\%,
and 17\,250\%. These comparisons show that the three-state response is
locally well represented by its Duhamel series, but a single expansion about
the zero-transfer model cannot be extrapolated across the full interval.
The separate zero-reference expansion and its error growth are documented
in Sec.~S3.3 of the Supporting Information.

\subsection{Interpretation}
\label{sec:VC}

The Duhamel expansion does more than provide a 
reconstruction algorithm --- it associates spectral 
changes with specific transport processes. Each 
order in the expansion corresponds to a physically 
distinct mechanism: first-order terms describe 
single pathway-mixing events, while second-order 
terms capture either nonlinear deformations of the 
generator or two successive mixing events separated 
by free evolution. Spectral deviations from the 
reference response are thereby decomposed into a 
hierarchy of transport processes rather than absorbed 
into phenomenological line-shape parameters.

\Rone{This decomposition suggests a natural inverse
problem. Given a reference Liouvillian and an observed
spectral response, one seeks the local Liouvillian
deformation operators required to account for the
measured change in pathway response. The present results
show both the promise and the limitation of performing
this reconstruction from derivatives evaluated at a
single reference point: surprisingly large spectral
deformations can be captured at low order, but
quantitative accuracy ultimately deteriorates as the
target model moves farther from the reference. This
observation motivates an incremental formulation in
which local response derivatives are reevaluated along
a trajectory through parameter space rather than
extrapolated from a single point.
}

\Rone{This limitation motivates the subsequent spectral
autodiff kernel expansion framework~\cite{bittner2026SAKE}
in which automatic differentiation supplies local response
derivatives along trajectories through model space.
Combined with the state--generator geometry developed in
Ref.~\citenum{Bittner:2026ab},
these local derivatives enable incremental transport of the
response between neighboring models. Rather than relying
on a single long-range expansion about a fixed reference
Liouvillian, the response can therefore be reconstructed
successively along a path
$\boldsymbol{\lambda}(s)$ through model space.}

\section{Discussion}
\label{sec:VI}

The results of this work demonstrate that basis
incompatibility between coherent and dissipative sectors
of an open-system generator produces systematic changes
in multidimensional spectroscopic response. In the
minimal model considered here, rotation of the
environment-selected dephasing basis relative to the
Hamiltonian basis introduces population--coherence mixing
during the reduced dynamics, producing characteristic
changes in spectral weight, line shape, and dispersive
structure. These effects provide a direct spectroscopic
signature of parametric deformation of the reduced
Liouvillian.

The Duhamel construction provides a local differential
description of this deformation. Derivatives of the
Liouvillian about a reference model determine the
corresponding derivatives of the finite-time propagator
and frequency-domain resolvents and, through them, the
derivatives of the nonlinear response. These local
deformation operators should not themselves be identified
with a geometric connection. A geometric connection
requires additional structure for comparing neighboring
physical models, such as that provided by the
state--generator geometry developed in
Ref.~\citenum{Bittner:2026ab}.

\Rtwo{The synthesis of these structures is the central
geometric content of the present formulation. The Duhamel
hierarchy supplies local response derivatives, while the
model-space connection specifies how such local information
is compared between neighboring points in parameter space.
Consequently, basis incompatibility need not be interpreted
as generating curvature by itself. Rather, it produces a
parameter-dependent deformation of the reduced dynamics
whose spectroscopic consequences can be transported and
compared covariantly on the underlying model manifold.}

\Rtwo{The minimal model isolates this response-deformation
mechanism in a controlled setting. Rotation of the
dephasing operator relative to the Hamiltonian basis
generates population--coherence mixing, while the
Hamiltonian and fixed amplitude-relaxation channel remain
unchanged. The model is not intended to represent a
specific microscopic environment, but rather to provide
a two-control family in which the relation between
Liouvillian deformation and nonlinear spectral response
can be examined directly. The main-text Taylor scan varies
$\theta$ at fixed $\gamma_\phi$, while the induced geometry
and full-domain convergence also resolve the dependence on
$\gamma_\phi$.}

The construction is formulated entirely at the level of
the reduced open quantum system. At the level of the
combined system and environment, the total dynamics may
be represented as unitary evolution on an enlarged
Hilbert space; after tracing over the environmental
degrees of freedom, their influence is encoded in the
effective reduced Liouvillian. Our objective is therefore
not to reconstruct individual bath modes, spectral
densities, or microscopic coupling mechanisms, but to
characterize how their effective action deforms the
reduced dynamics interrogated by nonlinear spectroscopy.
Different microscopic environments may consequently
produce similar spectroscopic deformations if they induce
equivalent changes in the effective Liouvillian.

The most striking numerical result is the range over
which this local construction remains informative.
The second-order Taylor/Duhamel expansion uses response
derivatives evaluated entirely at the aligned reference
point $\theta=0$, yet retains the dominant resonant and
dispersive spectral structure even for large pointer-basis
misalignment. These changes are not reducible to uniform
relaxation or environmental broadening: variations in
line shape, spectral weight, and dispersive structure
reflect the parameter-dependent mixing of population and
coherence sectors generated by the reduced Liouvillian.
At $\theta=\pi/2$, the global quantitative error has
grown to approximately $37\%$, but the principal spectral
features remain recognizable.

The additional convergence studies clarify the local range of this
interpretation. In the two-state model, average error decreases with order
over the full two-control domain, while the worst-case error remains large
near its edges; within the smaller neighborhood of the reference, the
second-order error remains below $0.96\%$ (Supporting Information,
Secs.~S2.2 and S2.3). The three-state results likewise show rapid
order-by-order improvement for a nearby reference and loss of convergence
when the same target is expanded from zero transfer (Supporting
Information, Secs.~S3.2 and S3.3).

The calculation therefore demonstrates both the
substantial information encoded in low-order local
response derivatives and the intrinsic limitation of
extrapolating from a single point in model space. The
Duhamel construction relates these spectral changes
directly to derivatives of the Liouvillian-dependent
propagators about the reference model, providing a
physically interpretable decomposition of the response
deformation rather than a phenomenological correction.

More generally, the information accessible through a
multidimensional spectrum depends on the observables used
to interrogate the reduced dynamics~\cite{paiva2026detection}.
Different measurement channels may therefore provide
complementary projections of the same underlying
Liouvillian deformation, an issue that becomes important
when considering the inverse reconstruction of open-system
dynamics from spectroscopic data.

\begin{acknowledgments}
CSA acknowledges funding from the Government of Canada (Canada Excellence Research Chair CERC-2022-00055),  
from the Institut Courtois, Facult\'e des arts et des sciences, Universit\'e de Montr\'eal (Chaire de recherche de direction de l'Institut Courtois) and from the Natural Science and Engineering Research Council of Canada (NSERC Discovery Grant RGPIN-2024-05893). 
ERB acknowledges funding from the National Science Foundation (CHE-2404788), Robert A.\ Welch Foundation (E-1337), and the U.S. Department of Energy, Office of Science, under Award No. DE-SC0025706. ERB gratefully acknowledges funding from IVADO for a Visiting Professorship at the Institut Courtois, Universit\'e de Montr\'eal. 
\end{acknowledgments}

\subsection*{Use of Generative Artificial Intelligence}
In compliance with institutional guidelines of the Universit\'e de Montr\'eal, generative artificial intelligence tools were used to assist with the editing of language and stylistic refinement of parts of the manuscript and to assist in the synthesis of the literature. These tools were not used to generate scientific content, perform analysis, or influence the interpretation of results. All content has been reviewed and validated by the authors, who assume full responsibility for the manuscript.

\section*{Data availability}

The numerical data and code that support the findings of this study are
available in the repository's \texttt{data\_availability/} directory and in
the Zenodo archive with DOI
\href{https://doi.org/10.5281/zenodo.24347090}{10.5281/zenodo.24347090}.
The repository package includes a Jupyter notebook and scripts for regenerating
the manuscript and Supporting Information figures and convergence tables,
together with numerical arrays and run summaries.



\bibliographystyle{apsrev4-2}
\bibliography{Refs-Local-clean}

\end{document}